\documentclass[aps,prb,twocolumn,groupedaddress]{revtex4}

\newif\ifpdf
\ifx\pdfoutput\undefined
\pdffalse 
\else
\pdfoutput=1 
\pdftrue
\fi

\ifpdf
\usepackage[pdftex]{graphicx}
\else
\usepackage{graphicx}
\fi
\usepackage{hyperref}

\begin{document}

\ifpdf
\DeclareGraphicsExtensions{.pdf , .jpg}
\else
\DeclareGraphicsExtensions{.eps, .jpg}
\fi

\def\hslash{\hbar}
\def\imag{i}
\def\grad{\vec{\nabla}}
\def\div{\vec{\nabla}\cdot}
\def\curl{\vec{\nabla}\times}
\def\DDt{\frac{d}{dt}}
\def\ddt{\frac{\partial}{\partial t}}
\def\ddx{\frac{\partial}{\partial x}}
\def\ddy{\frac{\partial}{\partial y}}
\def\lap{\nabla^{2}}
\def\divv{\vec{\nabla}\cdot\vec{v}}
\def\gradS{\vec{\nabla}S}
\def\vvec{\vec{v}}
\def\wc{\omega_{c}}
\def\<{\langle}
\def\>{\rangle}
\def\Tr{{\rm Tr}}
\def\Csch{{\rm csch}}
\def\Coth{{\rm coth}}
\def\Tanh{{\rm tanh}}
\def\g2{g^{(2)}}



\title{An orbital-free self-consistent field approach for molecular clusters and liquids}


\author{Sean W. Derrickson}
\affiliation{Department of Chemistry and Center for Materials Chemistry\\
University of Houston, Houston TX 77204}
\author{Eric R. Bittner}
\affiliation{Department of Chemistry and Center for Materials Chemistry\\
University of Houston, Houston TX 77204}

\date{\today}

\begin{abstract}
We present an ``orbital'' free density functional theory for computing 
the quantum ground state of atomic clusters and liquids.  Our approach 
combines the Bohm hydrodynamical description of quantum mechanics 
with an information theoretical approach to determine an optimal 
quantum density function in terms of density approximates to a
statistical sample.   
The ideas of Bayesian statistical
analysis and an expectation-maximization procedure are combined to
develop approximations to the quantum density  and
thus find the approximate quantum force.  The quantum force is then
combined with a Lennard-Jones force to simulate clusters of Argon
atoms and to obtain the ground state configurations and energies.
 As demonstration of the utility and flexibility of the approach, 
we compute the lowest energy structures for small rare-glass clusters.
Extensions to many atom systems is straightforward. 
\end{abstract}

\maketitle

%

\section{\label{intro}Introduction}

It has been long recognized that computational effort of grid-based
quantum mechanical methods for nuclear dynamical problems grows
exponentially with the number of degrees of freedom.  This limits the
size of systems that can be handled in an exact manner to those with 4
or less atoms. This is perhaps best illustrated in the field of
reactive scattering experiments which have been limited to systems
with 6D \cite{zhang:1146,zhang:4544,somers:11379,echave:402}, but it
is also clearly seen in other areas such as photodissociation
processes. \cite{guo:6562,manth:111} In light of this, considerable
progress has been made in developing rigorous approaches for
contracting the basis size required to perform such calculations.  One
such approach that has seen considerable success is the
multi-configurational time-dependent Hartree approach (MCTDH)
developed by Meyer and co-workers\cite{manthe:111,manthe:3199} that
overcomes this limitation in a numerically exact way by expanding the
time-dependent wave function in terms of a number of time-dependent
configurations.
 $$
\Psi(t) = \sum_J A_J(t) \Phi_J(t),
$$
in which the single particle (or quasi-particle)
basis functions $\Phi_J(t)$ and the expansion coefficients are
are coupled by the MCTDH equations of motion.

For condensed phase systems and liquids path integral Monte Carlo
(PIMC) and centroid based molecular dyamics remain the method of
choice. They have been extremely successful in calculating a wide
variety of different thermodynamic properties of heavily quantum
systems. \cite{calvo:7312,rick:6658,neirotti:3990,chakravarty:10663}
Despite the success of PIMC approaches of late there are some inherent
difficulties it faces. For instance, at low temperatures the amount of
parameters that must be included can become prohibitive and lead to
slow convergence.

We present an approach which can model low temperature Lennard-Jones
 clusters with ease. The method developed herein develops along
 analogous lines to the MCTDH approach and can be best thought of as
 an ``orbital'' free approach since we work entirely at the level of
 the nuclear $N$-body density. We do this by first writing the
 configurational density $n(1\cdots N)$ that describes the statistical
 likelihood of finding the system in a given multi-dimensional
 configuration $\{r_1\cdots r_N\}$ as a superposition of statistical
 approximates $p(r_1\cdots r_N,c_m)$ that are joint probabilities for
 finding system at $\{r_1\cdots r_N\}$ and that it is a variant of
 some statistical distribution described by the approximate. These
 approximates can be any elementary probability distribution function
 that can be specified in terms of its statistical moments, $c_m$, the
 simplest of which for our purposes are multi-dimensional gaussians.
 In this case, we need to be able to specify $m(3N(3N+1)/2 + 3N + 1) =
 {\cal O}(mN^2)$ variables corresponding to the elements of the
 covariance matrix, the central mean, and amplitude for $m$
 $3N$-dimensional gaussians.

Explicit correlations between various degrees of freedom can be
excluded in straightforward way by factoring the approximates.  For
example, if we factorize the full covariance matrix into individual
atomic components, the configurational density can be cast in terms of
the individual atomic densities
\begin{equation}
n(1\cdots N) = \prod_{i=1}^Nn_i(r_i).     \label{eq:densityapprox}
\end{equation}
We can then expand each atomic density $n_i(i)$ as a linear combination of density approximates. 
\begin{equation}
n_i(r_i) = \sum_{m=1}^M p_{mi}(c_{mi}, r_i),  
 \label{eq:approximates}
\end{equation}
This dramatically reduces the number of coefficients we need to
determine to $mN\times (6+3+1) = {\cal O}(mN)$.  Intermediate
factorization scheme yield similar scaling behavior allowing one to
tune the computational complexity of the system depending upon the
degree of correlation required by a particular physical problem.  For
example, one can define {\em quasi-atoms} by explicitly including
covariance between the degrees of freedom of 2 or more atoms.  As we
shall derive next, each quasi-atom or atom will then {\em evolve} in
the mean-field of the other quasi-atoms of the system.

In this paper, we present a grid-free adaptive hydrodynamic approach
for computing the quantum ground-state density for a system of $N$
nuclei.  Our approach uses Bayesian analysis to deduce from a
statistical sampling of the density the best set of $m$ statistical
approximates describing that density.  We then use a quantum
hydrodynamical scheme to move the sample points towards a minimal
energy configuration.  As proof of concept we consider the zero-point
energy of small 4 and 5 atom clusters of Argon and Neon with pair-wise
interatomic potential interactions.  Finally, we discuss how the
approach may be used to develop new quantum-classical and fully
quantum mechanical approaches for treating quantum mechanical solute
particles (such as an excess e$^-$ or He atom) in a liquid of
classical or quasi-classical atoms (such as Ar or Ne).

\section{Self-consistent field equations}

We begin by writing the full many-body Hamiltonian for the nuclear
motion of a collection of atoms with pair-wise interaction potentials.
\begin{eqnarray}
H = -\sum_{i=1}^N \frac{1}{2m_i}\nabla_i^2  + \sum_{i\ne j}V(ij),
\end{eqnarray}
where the first is the sum of the kinetic energies of the individual
 atoms and the second is the sum of the potential energy
 contributions.  For an arbitrary $N$-body trial density, the energy
 functional is given by
\begin{eqnarray}
E[n] = T[n] +  \sum_{i\ne j} \int\int n_i(r_i) n_j(r_j) V(ij) dr_i dr_j.
\end{eqnarray}
Since the kinetic energy operator is separable and we have assumed
distinguishability amongst the constituent atoms, the kinetic energy
term is the sum of the individual kinetic energy functionals.
\begin{eqnarray}
T[n(1\cdots N)] = \sum_{i=1}^N T_i[n(r_i)],
\end{eqnarray}
As in electronic structure DFT, evaluating the kinetic energy
functionals in an orbital free form is problematic since evaluating
the quantum kinetic energy operator is a non-local operator and the
density is a local function. \cite{parr:1112},

If instead we write the quantum wave function in polar form, as in the
 hydrodynamic formulation of quantum
 mechanics\cite{madelung:111,debroglie:111,debroglie:222},
\begin{eqnarray}
\Psi(1\cdots N)  = \sqrt{n(1\cdots N)}e^{i \phi(1\cdots N)},
\end{eqnarray}
we can arrive at a stationary condition that if $\grad\phi = 0$\cite{holland:111}, 
\begin{eqnarray}
V(1\cdots N)   - \sum_{i}\frac{1}{2m_i}\frac{1}{\sqrt{n(r_i)}}\nabla_i^2\sqrt{n(r_i)} = const ,    \label{eq:stationary}
\end{eqnarray}
at all points in space.  The constant is of course the energy.  By inspection, then, 
we can define our kinetic energy functional as 
\begin{eqnarray}
T[n(r_i)] = -\frac{1}{2m_i}\int \sqrt{n(r_i)}\nabla_i^2\sqrt{n(r_i)} dr_i.
\end{eqnarray}
Integrating by parts and letting $n(i)\rightarrow 0 $ at $\pm\infty$ produces the 
familiar von Weizsacker kinetic energy functional\cite{weizsacker:111}
\begin{eqnarray}
T_W[n(r_i)] = +\frac{1}{8m}\int \frac{1}{n(r_{i})}\grad_i n(r_{i})\cdot \grad_i n(r_i) dr_i.    \label{weizeq}
\end{eqnarray}
Thus, the total energy functional is given in terms of the single particle densities.
\begin{eqnarray}
E[n] = \sum_{i=1}^N T_W[n_i(r_i)] + \sum_{i\ne j} \int\int n_i(r_i) n_i(r_j) V(ij) dr_idr_j.
\end{eqnarray}
Taking the variation of $E[n]$ with respect to the single-particle densities 
with the constraint that $\sum_i \int n_i(r_i) di = N$,
\begin{widetext}
\begin{eqnarray}
\delta \left\{ \sum_{i=1}^N \left(T_W[n_i(r_i)] + \sum_{j\ne i} \int\int n_i(r_i) n_j(r_j) V(ij)dr_i dr_j -
\mu\left( \int n_i(r_i) dr_i - 1\right)\right)\right\} = 0,
\end{eqnarray}
\end{widetext}
leads to the following Euler-Lagrange equations:
\begin{eqnarray}
\frac{\delta T_W[n_i(r_i)]}{\delta n_i(r_i)} + \sum_{j\ne i}\int V(ij)n_j(r_j) dr_j -\mu = 0.
\end{eqnarray}
When satisfied, $\mu$ is the vibrational ground-state energy and the $n_i(r_i) = |\phi_i(i)|^2$
are the probability densities of the individual nuclei. 

Let us take a simple pedagogic case of a particle in a harmonic well,  taking
the trial density density to be a Gaussian, $n(x) = \sqrt{a/\pi}\exp(-a x^2)$.
Evaluating the energy functional yields: 
$$
E[n(x)]= \frac{1}{4 m}a + \frac{m\omega^2}{4 a}. 
$$
Minimizing with respect to the trial density
$$
\frac{\delta E[n]}{\delta n} = \frac{d E}{da} = 0,
$$
yields the familiar $E = \omega/2$ and $a = m\omega$. 
This idea is easy to extend beyond purely harmonic systems and 
gaussian trial functions.   Since $n(i)$ is a probability distribution
function, it is a  positive, real, and integrable function.  

In the next section, we show how the single particle densities can be
 estimated as super-positions of single particle density approximates
 and that the coefficients (rather moments) of the approximates can be
 optimized to compute both the ground state energy and the
 single-particle densities.

\section{Mixture Modeling \label{section:mm}}
The single-particle probability distribution 
functions (PDF) can be represented by a mixture
model \cite{gershenfeld:111,mclachlan:111} by summing a
finite number $M$ of density approximates
\begin{equation}
n (r) = \sum_{m}^{M} p({\bf r},c_{m}),       \label{eq:mixdens} 
\end{equation}
where $p({\bf r},cm)$ is the probability that a randomly chosen member
of the ensemble has the configuration ${\bf r}$ and is a variant of
the $m$th  approximate  designated by $c_{m}$. 
These approximates may be Gaussians or any other 
integrable multi-dimensional function which can be parameterized by its moments. 
For gaussian clusters, we have a weight $p(c_m)$, a mean position vector
$\mu_{m}$, and a covariance matrix $C_{m}$.  

By definition, each joint probability in Eq. \ref{eq:mixdens} is
related to a pair of conditional probabilities according to the
relation
\begin{equation}
p({\bf r}, c_{m}) = p(c_{m}) p({\bf r} |c_{m}) = n({\bf r}) p(c_{m}| {\bf r}),     \label{eq:condprob} 
\end{equation}
where the forward conditional probability $p({\bf r} | c_{m})$ refers
to the probability that a randomly chosen variant of $c_{m}$ has the
configuration $r$. Conversely, the posterior probability $p(c_{m}|
{\bf r})$ refers to the probability that the configuration point $r$
is a variant of the approximate $c_{m}$. In probability theory, $n(r)$
and $p(c_{m})$ are known as marginal probabilities; however, we shall
simply refer to them as the quantum density and weight of the $m$th
approximate, respectively. The expansion weights are strictly positive
semidefinite and sum to unity. Substituting the first equality of
Eq. \ref{eq:condprob} into Eq. \ref{eq:mixdens} we have
\begin{equation}
n({\bf r}) = \sum_{m}^{M} p(c_{m}) p({\bf r} | c_{m}).                   \label{eq:denssum} 
\end{equation}

We have considerable freedom at this point in specifying the 
exact functional form of the conditional probabilities as well as 
the degree of correlation within each conditional.  This freedom of specification 
allows us to construct ``models'' 
that explicitly take into account nonseparable
correlations in configuration space.  
For the case of gaussian approximates this is accomplished by 
keeping or discarding various off-diagonal terms incorporates in the 
covariance matrix, {\bf C},
\begin{equation}
p({\bf r} | c_{m}) = \sqrt{\frac{\| {\bf C}^{-1} \|}{(2\pi)^{N_{d}} }}
e^{({\bf r}_{d} - \mu_{m,d}).\bf{ C}_{m}^{-1}.({\bf r}_{d} -
\mu_{m,d})}. \label{eq:clusterwcov}
\end{equation}
The term,$\| {\bf C}^{-1} \|$, stands for the reciprocal of the
greatest value of the determinant of the covariance matrix. It is also
possible to construct a model that assumes that each approximate is
completely separable and takes the form of a product over the
$N_{d}$-dimensional configuration space, that reduces the covariance
matrix, $\bf{ C}$, to a variance vector,
\begin{equation}
p({\bf r} | c_{m}) = \prod_{d}^{N_{d}} \sqrt{\frac{1}{2\pi
\sigma_{m,d}^{2} }} \exp(({\bf r}_{d} - \mu_{m,d})^{2} /
(2\sigma_{m,d}^{2}) ). \label{eq:clusterwocov}
\end{equation}
Numerical tests by Maddox and Bittner indicate that separable case is
computationally faster for high dimensional systems, but produces a
less accurate estimate of the quantum ground-state
energy.~\cite{maddox:6465} For larger systems the noncovariant case
can certianly be used to speed calculations. We will examine the case
where there is nonzero covariance between the three dimensions, but
the atomic degrees of freedom have zero overlap. For a discussion of
the strengths and weaknesses involved with mixture models, one is
referred to the Refs. \cite{maddox:6465,heller:2923}.  This
approximation provides a sufficient approximation to the density for
the calculation of the ground state.
 
Once a model is decided upon one must then determine the parameters,
 in this case the Gaussian parameters $p(c_{m})$, $\mu_{m}$, and
 $C_{m}$, for each approximate from the statistical points
 representing the density. For instance the mean position vectors of
 the approximates are defined by the moments of the forward
 conditional probabilities
\begin{equation}
{\bf \mu}_{m} = \int {\bf r} p({\bf r} | c_{m}) d{\bf r}.         \label{eq:mufirst} 
\end{equation}
Rearranging Eq. \ref{eq:condprob} and substituting into
Eq. \ref{eq:mufirst}, we can write these parameters as
\begin{equation}
{\bf \mu}_{m} = \int {\bf r} \frac{n({\bf r}) p(c_{m} | {\bf r})
}{p(c_{m}) } d{\bf r},
\end{equation}
this is easily approximated by summing over an ensemble of points
$\{{\bf r}_n\}$ sampled from the $n(r)$ PDF,
\begin{equation}
{\bf \mu}_{m} \approx \frac{1}{N p(c_{m})} \sum_{n}^{N} {\bf r}_{n} p(c_{m} | {\bf r}_{n}).  \label{eq:musuming}
\end{equation}
We also define similar expressions for the covariance matrix and the expansion weights
\begin{equation}
p(c_{m}) \approx \frac{1}{N} \sum_{n}^{N} p(c_{m} | {\bf r}_{n}),
\label{eq:psuming}
\end{equation}
\begin{equation}
{\bf C}_{m} \approx \frac{1}{N p(c_{m})} \sum_{n}^{N} ( {\bf r}_{n} -
{\bf \mu})^{T} ( {\bf r}_{n} - {\bf \mu}) p( c_{m} | {\bf r}_{n}).
\label{eq:covsum}
\end{equation}
For the separable case, the variances are given by the diagonal
elements $\sigma_{m,i}^{2}= (C_{m})_{ii}$. The posterior terms
$p(c_{m}|{\bf r}_{n})$ for each ${\bf r}_n$ sample point in
Eqs. \ref{eq:musuming}-\ref{eq:covsum} are evaluated directly from the
forward probabilities according to Bayes' equation,
\begin{equation}
p(c_{m} | {\bf r}_{n}) = \frac{p(c_{m}) p({\bf r}_{n} | c_{m}) }
{\sum_{m} p(c_{m}) p({\bf r}_{n} | c_{m})}.  \label{eq:pospsum}
\end{equation}

Within this viewpoint, the sample points can be considered to be a
data set that represents the results of a series of successive
measurements.  Each data point carries an equal amount of information
describing the underlying quantum probability distribution
function. Bayes' equation gives the ratio of how well a given estimate
describes ${\bf r}_n$ to how well ${\bf r}_n$ is described by all of
the approximates. Thus, it represents the fraction of explanatory
information that a given sample point gives to the $m$-th
approximate. The estimate which {\em best } describes the point will
have the largest posterior probability at that
point. Eqs. \ref{eq:musuming}-\ref{eq:pospsum} can be iterated
self-consistently in order to determine the best possible set of
parameters that describe $n(r)$ in terms of a given ensemble of data
points. In doing so, we effectively maximize the log-likelihood that
the overall density model describes the entire collection of data
points.
$$
L = \log \prod_n n({\bf r}_n),
$$ Taking the variation of $L$ with respect to the model parameters
generates a series of update-rules for moving the approximates through
parameter space in the direction along
$\grad_{c_m}L$.\cite{gershenfeld:111}. For the case of Gaussian
approximates, the update rules for the mean, covariance matrix, and
marginal probabilities are given by,
$$
\delta \mu_{m}  =  \frac{{\bf C}_{m}}{N p(c_m)}  \grad_{\mu_m}L,
$$
$$
\delta {\bf C}_m = \frac{2( {\bf C}_m \otimes {\bf C}_m )}{ N p(c_m)}  \grad_{{\bf C}_m}L,
$$
$$
\delta p(c_m) = \frac{1}{N} ({\rm diag}[\Omega] - \Omega(\Omega)^{T})  \grad_{p(c_m)}L.
$$
Where $\otimes$ is the Kronecker product, $\Omega$ is the vector of all expansion weights, $\Omega = [p(c_1), \ldots, p(c_m)]^{T}$, and ${\rm diag}[\Omega]$ is a matrix with the elements from $\Omega$ constituting the diagonal entries. \cite{xu:111} 

The expectation maximization algorithm described above allows us to
generate an approximate analytical functional form for the single
particle density via statistical sampling over an ensemble of points.
The next step is to adjust the single-particle densities themselves to
produce a lower total energy.  We do this by deriving the quantum
hydrodynamic equations of motion for the sample points.
%
The quantum Hamilton-Jacobi equation generates the equations of motion
for the ray-lines of a time-dependent solution to the Schr\"odinger
equation.\cite{bohm:111222,bohm:111333,bohm:111444,wyatt:111}.
\begin{widetext}
\begin{eqnarray}
\frac{\partial S}{\partial t} + \sum_i\frac{|\grad_i S|^2}{2m_i} +\sum_{i\ne j} \int\int n_i(r_i) n_j(r_j) V(ij)dr_i dr_j 
 -\sum_i\frac{1}{2m_i}\frac{1}{\sqrt{n_i(r_i)}}\nabla_i^2 \sqrt{n_i(r_i)} = 0
\end{eqnarray}
 \end{widetext}
Since the density is separable into components, we easily arrive at a
set of time-dependent self-consistent field equations whereby the
motion of atom $i$ is determined by the average potential interaction
between atom $i$ and the rest of the atoms in the system.
\begin{eqnarray}
\dot S(i) &+& \frac{|\grad_i S|^2}{2m_i} + \sum_{j\ne i}\int V(ij) n_j(r_j) dr_j \nonumber \\
 &-&\frac{1}{2m_i}\frac{1}{\sqrt{n_i(r_i)}}\nabla_i^2 \sqrt{n_i(r_i)} = 0. \label{qhj}
\end{eqnarray}
Taking $\grad S = {\bf p} = m_i \dot {\bf r} $ as a momentum of a particle,
the equations of motion along a given ray-line or characteristic ${\bf r_n}(t)$ 
of the quantum wave function are given by 
\begin{eqnarray}
m_i\ddot{\bf r}_n =  -\sum_{j\ne i}\int (\grad_iV(ij)) n_j(r_j) dr_j  - \grad_i Q[n_i(r_i)]
\end{eqnarray}
where $Q[n(i)]$ is the Bohmian quantum potential specified by the last
term in Eq.~\ref{qhj}.  Stationary solutions of time-dependent
Schr{\"o}dinger equation are obtained whenever $m_i\ddot{\bf r}_n =
0$.  Consequently, by relaxing the sample points in a direction along
the energy gradient specified by
\begin{eqnarray}
\grad_iE= -\sum_{j\ne i}\int (\grad_iV(ij)) n_j(r_j) dr_j  - \grad_i Q[n_i(r_i)].\label{gradE}
\end{eqnarray}
keeping $n(r_j)$ fixed.  This generates a {\em new} statistical sampling, which we then use to 
determine a new set approximates. 

This process is similar to the semiclassical approximation strategy
for including quantum effects into otherwise classical calculations
introduced by Garaschuk and Rassolov
\cite{garashchuk:2482,garashchuk:111}. This semiclassical approximate
methodology is based upon de Broglie-Bohm trajectories and involves
the convolution of the quantum density with a minimum uncertainty wave
packet which is then expanded in a linear combination of Gaussian
functions
\begin{equation}
\rho(x) \approx f(x) = \sum_{n} c_{n}^{2} \exp[-a_{n}^{2}(x-X_{n})].      \label{lineargaus}
\end{equation}        
The Gaussian parameters $s=\{c_{n},X_{n},a_{n} \}$ in Eq. \ref{lineargaus} are determined by minimizing the functional 
\begin{equation}
F = \int [\rho(x) - f(x)]^{2} dx
\end{equation}
using an iterative procedure which explicitly involves solving the set
of equations $\partial F/\partial s_{k} = 0$. The parametrized density
leads to an approximate quantum potential (AQP) that is used to
propagate an ensemble of trajectories.  This approach has been used
successfully in computing reactive scattering cross-sections for the
co-linear H+H$_2$ reaction in one dimension.

It is important at this point to recognize the numerical difficulties
our group and others have faced in developing hydrodynamic trajectory
based approaches for time-dependent
systems. \cite{Lopreore:5190,wyatt:111,kendrick:5805,maddox:6465,bittner:1111,hugh:112}
The foremost difficulty is in the accurate evaluation of the quantum
potential from an irregular mesh of
points. \cite{Lopreore:5190,bittner:1111} The quantum potential is a
function of the local curvature of the density and can become singular
and rapidly varying as nodes form in the wave function or when the
wave function is sharply peaked, i.e. when $n^{1/2} \to 0$ faster than
$\nabla^2 n^{1/2} \to 0$. These inherent properties make an accurate
numerical calculation of the quantum potential and its derivatives
very difficult for all but the simplest systems.  These difficulties
are avioded in the cluster model approximation of the density, using
the expectation maximization (EM) algorithm,\cite{maddox:6465}.  By
obtaining a global optimal function that describes the density, we can
{\em analytically} compute the quantum force with great accuracy.  The
issue of nodes is essentially avoided so long as we are judicious in
our choice of density approximates.  If we choose node-free
approximates, then our overall density will likewise be node free.
For the purpose of determining vibrational ground-states, this seems
to be a worthwhile compromise.

The algorithm can be summarized as follows
\begin{enumerate}
\item For each atom, generate and sample a normalized trial density
$n_i(r_i)$.
\item Using the EM routines and the given sample of points, compute
the coefficients for the density approximates.
\item Compute the forces on each point using Eq.~\ref{gradE} and
advance each point along the energy gradient for one ``time'' step,
either discarding or dampening the velocity of each point.  This
generates a new sample of points describing the single-particle
density for each atom.  The new distribution should have a lower total
energy since we moved the sample points in the direction towards lower
energy.
\end{enumerate}
 Iterating through these last two steps,  we rapidly converge towards the energy minimum of the system.  

\section{Vibrational ground state of rare-gas clusters \label{section:results}}

As proof of concept we examine the ground vibrational state energies
of Lennard-Jones clusters.  The simple Lennard-Jones pair potential
provides reliable information about complicated systems, and has been
used in a number of recent studies with just a few examples listed in
Refs. \cite{berry:111,lynden-bell:1460,predescu:154305}.
Ground state energies of rare gas clusters are easily enough modeled
with molecular dynamics simulations, but the quantum corrections are
often quite large for these cases. These corrections are important
because the quantum character strongly affects the thermodynamics via
changes in the ground state structure due to increasing zero-point
energies \cite{calvo:7312}.  The zero point energy corrections for the
small clusters modeled here can be up to $0.66$kJ/mole. Indeed,
quantum corrections have been shown to lower solid to liquid
transition temperatures by approximately $10\%$, and the zero point
energy for small clusters can account for up to $35\%$ of the
classical binding energy.\cite{chakravarty:956}.

The effects from quantum delocalization are intuitively understood in
 the present approach through the quantum potential term in the
 equations of motion.  This explains why the quantum delocalization
 can account for a lowering of the transition temperature because some
 kinetic energy is always present even at very low temperatures. This
 spreading of the wave packet is known as a ``softening" of the
 crystal which leads to a lowering of the melting
 temperature.\cite{chakravarty:8938} These effects have been studied
 in the context of the transition from molecular to bulk-like
 properties of clusters.

In the calculations presented here, we used 300 statistical points to 
represent the density of each atom and propagated the SCF equations described above 
until the energy and the density were sufficiently converged.  Typically,
 this required 1.5 million to 3 million cycles.  Along the course of the energy
minimization, we strongly damped the time-evolution of the sample points to
 eliminate as much of the oscillations and breathing of the 
density components as possible.  

The Lennard-Jones parameters for the Argon clusters are 
$\epsilon = 0.9976$ kJ/mole and $\sigma = 3.42\AA$, and
 $\epsilon = 0.3059$ kJ/mole and $\sigma = 2.79\AA$ for the
  Neon clusters \cite{livesly:111}. Initial configurations for the 
  simulations are chosen to be close to the classical molecular
   dynamics minimum energy geometry, and are given some initial 
   Gaussian spread. 
 
\begin{figure}
\includegraphics[width=\columnwidth]{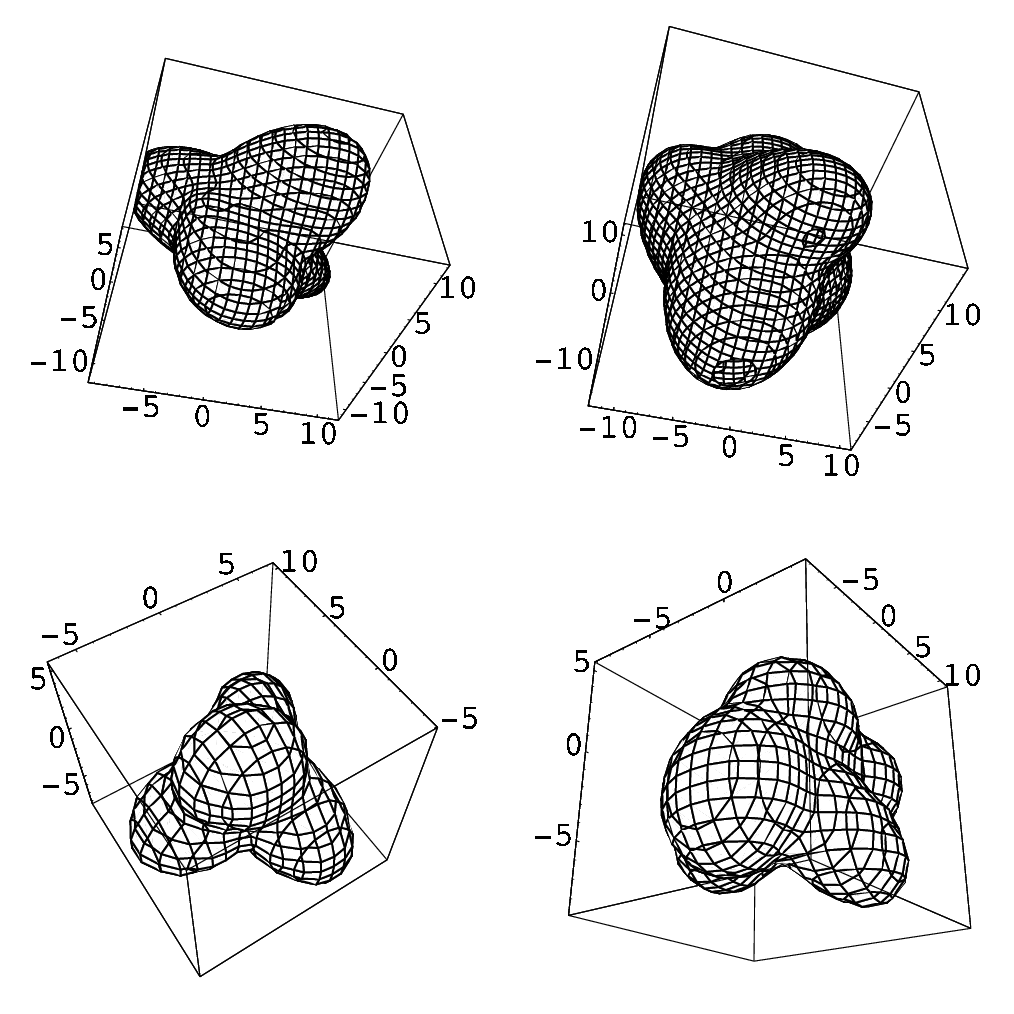}%
\caption{The isodensity contour plots of the clusters at a value of
$0.006$. In the upper left is the Ar$_{4}$ cluster, in the upper right
is the Ne$_{4}$, lower left has the Ar$_{5}$, and then bottom right is
Ne$_{5}$. The axis are listed in atomic units. \label{all4contour}}
\end{figure}

We show in Fig. \ref{all4contour} isodensity (0.006) contour plots for
the Ar$_4$, Ar$_5$, Ne$_4$, and
Ne$_5$.\footnote{http://eiger.chem.uh.edu:8080/webMathematica/pchem-apps/Plot3DLive.jsp}.
One can see quite clearly the underlying three-dimensional shape of
the cluster along with the delocalization of each atom about its
central location.  Each density "lobe" is nearly spherical with some
elongation. These density plots give a suggestive view of the overlap
of the densities which is ignored in Eq. \ref{eq:densityapprox} and
therefor ultimately in Eq. \ref{weizeq}. For this system this overlap
turns out to be minor, but for atoms such as Helium this would have to
be taken into account.

 \begin{figure}[t]
 \includegraphics[width=0.45\columnwidth]{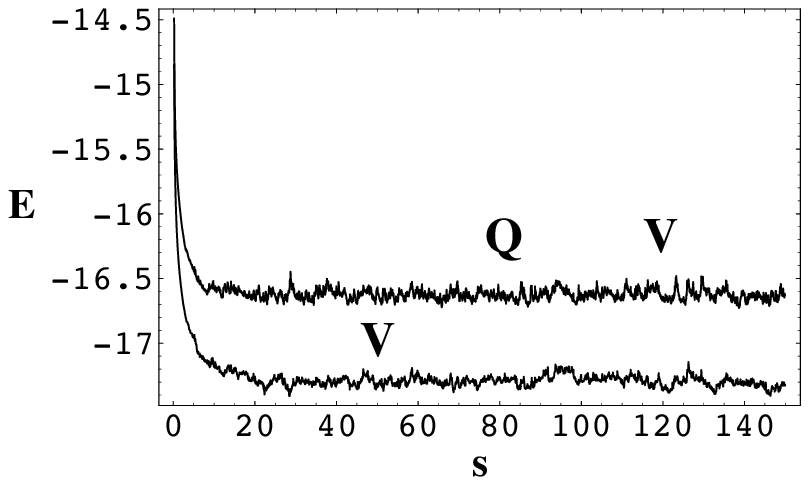}%
 \includegraphics[width=0.45\columnwidth]{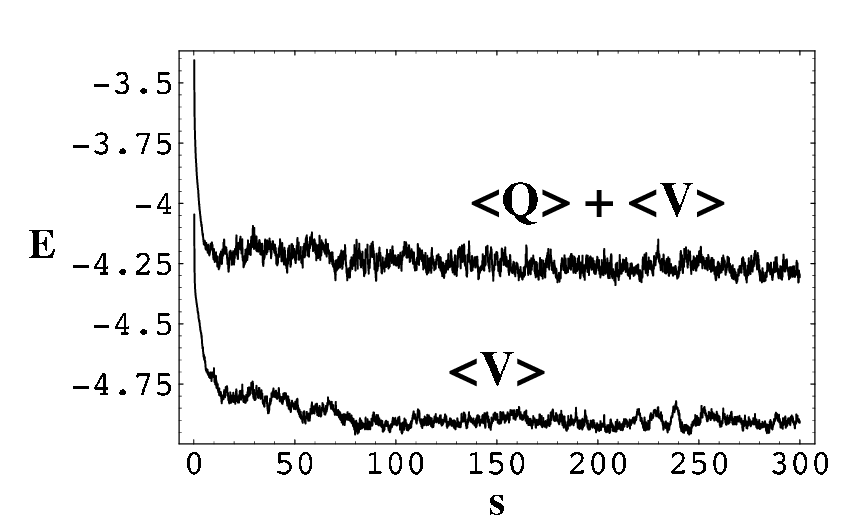}%
 \caption{The average potential energy $\langle V\rangle$ and total energy  $\langle Q\rangle + \langle V\rangle$ 
 of the Ar$_{5}$ and Ne$_5$ clusters in kJ/mole. The steps are measured in millions.\label{ar5bothener}}
 \end{figure}

In Fig.~\ref{ar5bothener} we show the total energy an the total
 potential energy for the Ar$_5$ and Ne$_5$ clusters as the system
 converges towards its lowest energy state.  Initially there is a
 rapid restructuring of the densities as they adjust to find a close
 approximation to the actual ground state density.  Following this
 initial rapid convergence, there is slower convergence phase as the
 density is further refined.  During this process as the sample points
 look for a configuration which fully equalizes of the quantum and
 kinetic energy terms from Eq. \ref{eq:stationary} the density
 approximation can sometimes prove inadequate and points can be pushed
 into temporary higher energy regions.  This leads to the fluctuations
 seen in the energies and any other averaged quantity, such as the
 interatomic distances.  In order to compute meaningful values for the
 energy and distances, we averaged these quantities over the last half
 million or so cycles. As can be seen from Figure \ref{ar5bothener}
 the Ne$_5$ cluster is slower to converge, but eventually does so
 around step 200(million).
 
 Tables \ref{table:fivea} and \ref{table:foura} lists the averaged
 interatom distances for each cluster compared to the equilibrium
 distances for the corresponding classical case.  For the case of
 $Ar_5$ the numerical fluctuations lead to an uncertainly of about
 0.3\% in the interatomic distances and for $Ne_5$, a 0.5\%
 uncertainty in the interatomic distances. it is important to note
 that the fluctuations mentioned here get smaller for larger systems
 as can seen by comparing the results for Ar$_5$ with Ar$_4$. This is
 because the well depths for the sample points become more
 pronounced. This has important implications since we hope to extent
 this method to larger systems.  Ne$_4$ can be seen to have the
 largest fluctuations, which is expected since it is the most quantum
 mechanical. All in all these values compare well with the classical
 distances.  In general, the quantum distances are slightly larger due
 to the fact that the gaussian atom densities are sampling part of the
 anharmonic attractive portion of the pair-potential.
 
  \begin{table}
 \caption{Inter-atomic distances for X$_5$ clusters in atomic units.\label{table:fivea}}
 \begin{ruledtabular}
 \begin{tabular}{ c | c | c | c | c }  
distances         & Argon          &    Argon (cl) & Neon    &  Neon (cl) \\  \hline 
$rd_{1,2}$       &    7.343$\pm$.019       &       7.225     & 6.167$\pm$.045   &   5.927    \\  \hline
$rd_{1,3}$       &    7.327$\pm$.018        &       7.225     &  6.115$\pm$.031  &   5.927   \\  \hline
$rd_{1,4}$       &    7.288$\pm$.018        &       7.199     &   6.057$\pm$.035  &   5.906 \\  \hline
$rd_{1,5}$       &    7.269$\pm$.016        &       7.199     &  6.048$\pm$.027   &   5.906    \\  \hline
$rd_{2,3}$       &    7.339$\pm$.023        &       7.225     &   6.162$\pm$.055  &   5.927     \\  \hline
$rd_{2,4}$       &    7.276$\pm$.018        &       7.199     &   6.060$\pm$.039   &   5.906     \\  \hline
$rd_{2,5}$       &    7.285$\pm$.016        &       7.199     &   6.070$\pm$.035  &   5.906     \\  \hline
$rd_{3,4}$       &    7.266$\pm$.018        &       7.199     &  6.112$\pm$.060     &   5.906     \\  \hline
$rd_{3,5}$       &    7.271$\pm$.016        &       7.199     &   6.077$\pm$.035  &   5.906     \\  \hline
$rd_{4,5}$     &     11.832$\pm$.020       &     11.7349  &   9.848$\pm$.039     &  9.627     
 \end{tabular}
 \end{ruledtabular}
 \end{table}

\begin{table}[h]
 \caption{Inter-atomic distances for X$_4$ clusters in atomic units. \label{table:foura}}
 \begin{ruledtabular}
 \begin{tabular}{ c | c | c | c | c }  
 distances        &   Argon   &  Argon (cl) & Neon &  Neon (cl)  \\  \hline 
$rd_{1,2}$       &    7.362$\pm$.024        &       7.21  &   6.167$\pm$ .043  &   5.918  \\  \hline
$rd_{1,3}$       &    7.296$\pm$.017        &       7.21  &   6.104 $\pm$ .051 &   5.918  \\  \hline
$rd_{1,4}$       &    7.294$\pm$.019        &       7.21  &   6.092 $\pm$ .049 &   5.918 \\  \hline
$rd_{2,3}$       &    7.320$\pm$.022        &       7.21  &   6.140 $\pm$ .038  &   5.918 \\  \hline
$rd_{2,4}$       &    7.305$\pm$.016        &       7.21  &   6.077 $\pm$ .027  &   5.918 \\  \hline
$rd_{3,4}$       &    7.286$\pm$.017        &       7.21  &   6.094 $\pm$ .040  &   5.918
 \end{tabular} 
 \end{ruledtabular}
 \end{table}

Table \ref{table:energies} summarizes the various contributions to the
the total energy for each cluster .  The ``classical" energies, $V_c$
are the energy minimum of the total potential energy surface
corresponding to the classical equilibrium configuration.  $\langle
V\rangle$, $\langle Q\rangle$, and $\langle E\rangle$ are the total
quantum potential energy, kinetic energy, and total energy of each
cluster.  The difference between the classical potential minimum,
V$_{c}$ and $\langle$E$\rangle$ is the zero point energy.  For our
results we calculate a virial-like term which is the ratio of the
kinetic to the potential energy of the system, $\langle Q\rangle /
\langle V \rangle$. It turns out the values we get for this term, seen
in Table \ref{table:energies}, are remarkably close to the values of
the de Boer parameter, $\Lambda$, for the atoms we are examining.
\begin{widetext}
\begin{table}[h]
 \caption{Table of the energies of the clusters in kJ/mole. \label{table:energies}}
 \begin{ruledtabular}
 \begin{tabular}{ c | c | c | c | c }  
         &     Ar$_{4}$ &Ne$_{4}$& Ar$_{5}$ & Ne$_{5}$ \\  \hline 
$V_{c}$  &    -11.972        &       -3.655  &   -18.166  &   -5.545  \\  \hline
    $\langle V \rangle$               &    -11.337$\pm$.225        &    -3.184$\pm$ .056     &  -17.302 $\pm$ .344   &   -4.895$\pm$.055  \\  \hline
     $\langle Q\rangle$               &    0.462$\pm$.033        &       0.460 $\pm$ .034 &   0.669 $\pm$ .041  &   0.646$\pm$.036  \\  \hline
     $\langle E \rangle$              &    -10.874$\pm$.215        &       -2.724 $\pm$ .049  &   -16.632 $\pm$ .330 &   -4.249$\pm$.043 \\  \hline
  $\langle Q\rangle/\langle V\rangle$         &     0.041       &       0.144  &   0.038  &   0.132 
 \end{tabular} 
 \end{ruledtabular}
 \end{table}
\end{widetext}
The de Boer parameter has been used in attempts to rationalize the
effects of quantum delocalization for Lennard-Jones systems.
\cite{chakravarty:956,chakravarty:10663,calvo:7312} Each Lennard-Jones
system can be defined in terms of its parameters $\epsilon$, or its
potential depth, its length scale of $\sigma$, and mass $m$. For a
given set of parameters, the thermal de Broglie wavelength, $\lambda =
\hbar /\sigma \sqrt{m k_{B} T}$ provides a means of approximating the
quantum effects, at some reduced temperature of the system, $T^{*} =
k_{B}T/\epsilon$. Further taking the ratio of $\lambda$ for two
different sets of parameters provides a means of comparing the quantum
effects of one system versus the other. This leads to the de Boer
parameter, $\Lambda = \hbar /r_{m} \sqrt{m \epsilon}$, which is the
ratio of the de Broglie wavelength of an atom with energy $\epsilon$,
with an intermolecular distance, $r_m$. Basically, the de Boer
parameter is useful for comparing the quantum character of different
Lennard-Jones clusters or liquids at a given temperature, in our case
zero. \cite{kubo:111} $\Lambda$ has a classical limit of $\Lambda = 0$
and anything above $\Lambda \approx 0.3$ is considered a quantum
system.  For Argon the de Boer parameter is, $\Lambda \approx 0.03$
which corresponds to a classical system, and quantum effects can be
treated as a perturbation. For Neon the de Boer parameter turns out to
be $\Lambda \approx 0.1$ which is a {\em quasi-quantum} system.

The de Boer parameter measures the delocalization of the system
compared to its size. The virial like term measures the percentage
energy contained in the kinetic term of the Hamiltonian. The kinetic
energy, also called the quantum potential energy, is also a measure of
the delocalization of the system. So we see that the two terms are
essentially measuring the same entity. For our results we also see a
possible trend to smaller values of $\langle Q\rangle / \langle V
\rangle$, as the system gets larger.

  


\section{Discussion}
A method for calculating ground state configuration of quantum
clusters and liquids has been outlined based upon some previous work
in approximating densities as quantum statistical distribution
function. The quantum and the Lennard-Jones potentials are used to
propagate an ensemble of Monte Carlo statistical points, in a DFT like
procedure. This is an orbital free approach in the sense that we only
work at the level of the nuclear density. In order to do this we
outline a ``cluster" model and expectation maximization (EM) algorithm
which is used to obtain the density in terms of the statistical points
representing each atom. The Lennard-Jones potential is calculated in a
mean field sense by averaging over the statistical points of each
atom, and the quantum potential is calculated from the density
obtained in the EM algorithm. Results were presented for 4 and 5 atom
clusters of Argon and Neon.  The results indicate good agreement with
the general classical results, but the quantum corrections can be seen
to be significant. Also shown is that the virial term we measure to
approximate the quantum effects is related to the de Boer parameter
used in previous studies.

 
The method outlined also seems to provide a means of artificial
control of the amount of quantum mechanical information desired from a
calculation. The covariances between atoms may be set to zero as we
have done, which corresponds to distinguishable particles, but keeping
the interatomic covariances seems to provide a possible path to
including other effects such as exchange energies and the like. This
all comes at the expense of increased complexity and reduced
computational speed.

One could also {\it tune} the amount of quantum effects by making
$\hbar$ a parameter that can vary between the values of $0$ to $1$, in
atomic units, in the equation for the quantum potential. Additionally
the method is easily extended to potentials beyond the
Lennard-Jones. For instance, covalent bonds could be modeled with
harmonic oscillators or Morse potentials, and Coulomb potentials are
easily modeled.

\begin{acknowledgments}
 The authors would like to thank Jeremy Maddox for helpful discussions. 
 This work was supported by grants from the National Science Foundation and 
 the Robert Welch Foundation. 
 \end{acknowledgments}


\begin{thebibliography}{40}
\expandafter\ifx\csname natexlab\endcsname\relax\def\natexlab#1{#1}\fi
\expandafter\ifx\csname bibnamefont\endcsname\relax
  \def\bibnamefont#1{#1}\fi
\expandafter\ifx\csname bibfnamefont\endcsname\relax
  \def\bibfnamefont#1{#1}\fi
\expandafter\ifx\csname citenamefont\endcsname\relax
  \def\citenamefont#1{#1}\fi
\expandafter\ifx\csname url\endcsname\relax
  \def\url#1{\texttt{#1}}\fi
\expandafter\ifx\csname urlprefix\endcsname\relax\def\urlprefix{URL }\fi
\providecommand{\bibinfo}[2]{#2}
\providecommand{\eprint}[2][]{\url{#2}}

\bibitem[{\citenamefont{Zhang and Zhang}(1994)}]{zhang:1146}
\bibinfo{author}{\bibfnamefont{D.~H.} \bibnamefont{Zhang}} \bibnamefont{and}
  \bibinfo{author}{\bibfnamefont{J.~Z.~H.} \bibnamefont{Zhang}},
  \bibinfo{journal}{The Journal of Chemical Physics}
  \textbf{\bibinfo{volume}{101}}, \bibinfo{pages}{1146} (\bibinfo{year}{1994}),
  \urlprefix\url{http://link.aip.org/link/?JCP/101/1146/1}.

\bibitem[{\citenamefont{Zhang and Light}(1996)}]{zhang:4544}
\bibinfo{author}{\bibfnamefont{D.~H.} \bibnamefont{Zhang}} \bibnamefont{and}
  \bibinfo{author}{\bibfnamefont{J.~C.} \bibnamefont{Light}},
  \bibinfo{journal}{The Journal of Chemical Physics}
  \textbf{\bibinfo{volume}{104}}, \bibinfo{pages}{4544} (\bibinfo{year}{1996}),
  \urlprefix\url{http://link.aip.org/link/?JCP/104/4544/1}.

\bibitem[{\citenamefont{Somers et~al.}(2004)\citenamefont{Somers, Olsen,
  Busnengo, Baerends, and Kroes}}]{somers:11379}
\bibinfo{author}{\bibfnamefont{M.~F.} \bibnamefont{Somers}},
  \bibinfo{author}{\bibfnamefont{R.~A.} \bibnamefont{Olsen}},
  \bibinfo{author}{\bibfnamefont{H.~F.} \bibnamefont{Busnengo}},
  \bibinfo{author}{\bibfnamefont{E.~J.} \bibnamefont{Baerends}},
  \bibnamefont{and} \bibinfo{author}{\bibfnamefont{G.~J.} \bibnamefont{Kroes}},
  \bibinfo{journal}{The Journal of Chemical Physics}
  \textbf{\bibinfo{volume}{121}}, \bibinfo{pages}{11379}
  (\bibinfo{year}{2004}),
  \urlprefix\url{http://link.aip.org/link/?JCP/121/11379/1}.

\bibitem[{\citenamefont{Echave and Clary}(1994)}]{echave:402}
\bibinfo{author}{\bibfnamefont{J.}~\bibnamefont{Echave}} \bibnamefont{and}
  \bibinfo{author}{\bibfnamefont{D.~C.} \bibnamefont{Clary}},
  \bibinfo{journal}{The Journal of Chemical Physics}
  \textbf{\bibinfo{volume}{100}}, \bibinfo{pages}{402} (\bibinfo{year}{1994}),
  \urlprefix\url{http://link.aip.org/link/?JCP/100/402/1}.

\bibitem[{\citenamefont{Guo et~al.}(1991)\citenamefont{Guo, Lao, Schatz, and
  Hammerich}}]{guo:6562}
\bibinfo{author}{\bibfnamefont{H.}~\bibnamefont{Guo}},
  \bibinfo{author}{\bibfnamefont{K.~Q.} \bibnamefont{Lao}},
  \bibinfo{author}{\bibfnamefont{G.~C.} \bibnamefont{Schatz}},
  \bibnamefont{and} \bibinfo{author}{\bibfnamefont{A.~D.}
  \bibnamefont{Hammerich}}, \bibinfo{journal}{The Journal of Chemical Physics}
  \textbf{\bibinfo{volume}{94}}, \bibinfo{pages}{6562} (\bibinfo{year}{1991}),
  \urlprefix\url{http://link.aip.org/link/?JCP/94/6562/1}.

\bibitem[{\citenamefont{Manth and Koppel}(1991)}]{manth:111}
\bibinfo{author}{\bibfnamefont{U.}~\bibnamefont{Manth}} \bibnamefont{and}
  \bibinfo{author}{\bibfnamefont{H.}~\bibnamefont{Koppel}},
  \bibinfo{journal}{Chem. Phys. Lett.} \textbf{\bibinfo{volume}{178}},
  \bibinfo{pages}{36} (\bibinfo{year}{1991}).

\bibitem[{\citenamefont{Manthe et~al.}(1990)\citenamefont{Manthe, Meyer, and
  Cederbaum}}]{manthe:111}
\bibinfo{author}{\bibfnamefont{U.}~\bibnamefont{Manthe}},
  \bibinfo{author}{\bibfnamefont{H.-D.} \bibnamefont{Meyer}}, \bibnamefont{and}
  \bibinfo{author}{\bibfnamefont{L.~S.} \bibnamefont{Cederbaum}},
  \bibinfo{journal}{Chemical Physics Letters} \textbf{\bibinfo{volume}{165}},
  \bibinfo{pages}{73} (\bibinfo{year}{1990}).

\bibitem[{\citenamefont{Manthe et~al.}(1992)\citenamefont{Manthe, Meyer, and
  Cederbaum}}]{manthe:3199}
\bibinfo{author}{\bibfnamefont{U.}~\bibnamefont{Manthe}},
  \bibinfo{author}{\bibfnamefont{H.-D.} \bibnamefont{Meyer}}, \bibnamefont{and}
  \bibinfo{author}{\bibfnamefont{L.~S.} \bibnamefont{Cederbaum}},
  \bibinfo{journal}{The Journal of Chemical Physics}
  \textbf{\bibinfo{volume}{97}}, \bibinfo{pages}{3199} (\bibinfo{year}{1992}),
  \urlprefix\url{http://link.aip.org/link/?JCP/97/3199/1}.

\bibitem[{\citenamefont{Calvo et~al.}(2001)\citenamefont{Calvo, Doye, and
  Wales}}]{calvo:7312}
\bibinfo{author}{\bibfnamefont{F.}~\bibnamefont{Calvo}},
  \bibinfo{author}{\bibfnamefont{J.~P.~K.} \bibnamefont{Doye}},
  \bibnamefont{and} \bibinfo{author}{\bibfnamefont{D.~J.} \bibnamefont{Wales}},
  \bibinfo{journal}{The Journal of Chemical Physics}
  \textbf{\bibinfo{volume}{114}}, \bibinfo{pages}{7312} (\bibinfo{year}{2001}),
  \urlprefix\url{http://link.aip.org/link/?JCP/114/7312/1}.

\bibitem[{\citenamefont{Rick et~al.}(1991)\citenamefont{Rick, Leitner, Doll,
  Freeman, and Frantz}}]{rick:6658}
\bibinfo{author}{\bibfnamefont{S.~W.} \bibnamefont{Rick}},
  \bibinfo{author}{\bibfnamefont{D.~L.} \bibnamefont{Leitner}},
  \bibinfo{author}{\bibfnamefont{J.~D.} \bibnamefont{Doll}},
  \bibinfo{author}{\bibfnamefont{D.~L.} \bibnamefont{Freeman}},
  \bibnamefont{and} \bibinfo{author}{\bibfnamefont{D.~D.}
  \bibnamefont{Frantz}}, \bibinfo{journal}{The Journal of Chemical Physics}
  \textbf{\bibinfo{volume}{95}}, \bibinfo{pages}{6658} (\bibinfo{year}{1991}),
  \urlprefix\url{http://link.aip.org/link/?JCP/95/6658/1}.

\bibitem[{\citenamefont{Neirotti et~al.}(2000)\citenamefont{Neirotti, Freeman,
  and Doll}}]{neirotti:3990}
\bibinfo{author}{\bibfnamefont{J.~P.} \bibnamefont{Neirotti}},
  \bibinfo{author}{\bibfnamefont{D.~L.} \bibnamefont{Freeman}},
  \bibnamefont{and} \bibinfo{author}{\bibfnamefont{J.~D.} \bibnamefont{Doll}},
  \bibinfo{journal}{The Journal of Chemical Physics}
  \textbf{\bibinfo{volume}{112}}, \bibinfo{pages}{3990} (\bibinfo{year}{2000}),
  \urlprefix\url{http://link.aip.org/link/?JCP/112/3990/1}.

\bibitem[{\citenamefont{Chakravarty}(1995{\natexlab{a}})}]{chakravarty:10663}
\bibinfo{author}{\bibfnamefont{C.}~\bibnamefont{Chakravarty}},
  \bibinfo{journal}{The Journal of Chemical Physics}
  \textbf{\bibinfo{volume}{103}}, \bibinfo{pages}{10663}
  (\bibinfo{year}{1995}{\natexlab{a}}),
  \urlprefix\url{http://link.aip.org/link/?JCP/103/10663/1}.

\bibitem[{\citenamefont{Parr and Yang}(1989)}]{parr:1112}
\bibinfo{author}{\bibfnamefont{R.~G.} \bibnamefont{Parr}} \bibnamefont{and}
  \bibinfo{author}{\bibfnamefont{W.}~\bibnamefont{Yang}},
  \emph{\bibinfo{title}{Density Functional Theory of Atoms and Molecules}}
  (\bibinfo{publisher}{Oxford [England] : Clarendon Press ; New York : Oxford
  University Press}, \bibinfo{year}{1989}).

\bibitem[{\citenamefont{Madelung}(1926)}]{madelung:111}
\bibinfo{author}{\bibfnamefont{E.}~\bibnamefont{Madelung}},
  \bibinfo{journal}{Z. Phys.} \textbf{\bibinfo{volume}{40}},
  \bibinfo{pages}{322} (\bibinfo{year}{1926}).

\bibitem[{\citenamefont{de~Broglie}(1926)}]{debroglie:111}
\bibinfo{author}{\bibfnamefont{L.}~\bibnamefont{de~Broglie}},
  \bibinfo{journal}{C. R. Acad. Sci. Paris} \textbf{\bibinfo{volume}{183}},
  \bibinfo{pages}{447} (\bibinfo{year}{1926}).

\bibitem[{\citenamefont{de~Broglie}(1927)}]{debroglie:222}
\bibinfo{author}{\bibfnamefont{L.}~\bibnamefont{de~Broglie}},
  \bibinfo{journal}{C. R. Acad. Sci. Paris} \textbf{\bibinfo{volume}{184}},
  \bibinfo{pages}{273} (\bibinfo{year}{1927}).

\bibitem[{\citenamefont{Holland}(1993)}]{holland:111}
\bibinfo{author}{\bibfnamefont{P.~R.} \bibnamefont{Holland}},
  \emph{\bibinfo{title}{The Quantum Theory of Motion}}
  (\bibinfo{publisher}{Cambridge University Press, New York},
  \bibinfo{year}{1993}).

\bibitem[{\citenamefont{von Weizsacker}(1935)}]{weizsacker:111}
\bibinfo{author}{\bibfnamefont{C.~F.} \bibnamefont{von Weizsacker}},
  \bibinfo{journal}{Z. Phys.} \textbf{\bibinfo{volume}{96}},
  \bibinfo{pages}{431} (\bibinfo{year}{1935}).

\bibitem[{\citenamefont{Gershenfeld}(1999)}]{gershenfeld:111}
\bibinfo{author}{\bibfnamefont{N.}~\bibnamefont{Gershenfeld}},
  \emph{\bibinfo{title}{The Nature of Mathematical Modeling}}
  (\bibinfo{publisher}{Cambridge University Press, Cambridge, U.K.},
  \bibinfo{year}{1999}).

\bibitem[{\citenamefont{McLachlan and Basford}(1998)}]{mclachlan:111}
\bibinfo{author}{\bibfnamefont{G.~J.} \bibnamefont{McLachlan}}
  \bibnamefont{and} \bibinfo{author}{\bibfnamefont{K.~E.}
  \bibnamefont{Basford}}, \emph{\bibinfo{title}{Mixture models: Inference and
  Applications to Clustering}} (\bibinfo{publisher}{Dekker, Inc., New York,
  1988}, \bibinfo{year}{1998}).

\bibitem[{\citenamefont{Maddox and Bittner}(2003)}]{maddox:6465}
\bibinfo{author}{\bibfnamefont{J.~B.} \bibnamefont{Maddox}} \bibnamefont{and}
  \bibinfo{author}{\bibfnamefont{E.~R.} \bibnamefont{Bittner}},
  \bibinfo{journal}{The Journal of Chemical Physics}
  \textbf{\bibinfo{volume}{119}}, \bibinfo{pages}{6465} (\bibinfo{year}{2003}),
  \urlprefix\url{http://link.aip.org/link/?JCP/119/6465/1}.

\bibitem[{\citenamefont{Heller}(1981)}]{heller:2923}
\bibinfo{author}{\bibfnamefont{E.~J.} \bibnamefont{Heller}},
  \bibinfo{journal}{The Journal of Chemical Physics}
  \textbf{\bibinfo{volume}{75}}, \bibinfo{pages}{2923} (\bibinfo{year}{1981}),
  \urlprefix\url{http://link.aip.org/link/?JCP/75/2923/1}.

\bibitem[{\citenamefont{Xu and Jordan}(1996)}]{xu:111}
\bibinfo{author}{\bibfnamefont{L.}~\bibnamefont{Xu}} \bibnamefont{and}
  \bibinfo{author}{\bibfnamefont{M.~I.} \bibnamefont{Jordan}},
  \bibinfo{journal}{Neural Computation} \textbf{\bibinfo{volume}{8}},
  \bibinfo{pages}{129} (\bibinfo{year}{1996}).

\bibitem[{\citenamefont{Bohm}(1952{\natexlab{a}})}]{bohm:111222}
\bibinfo{author}{\bibfnamefont{D.}~\bibnamefont{Bohm}}, \bibinfo{journal}{Phys.
  Rev.} \textbf{\bibinfo{volume}{85}}, \bibinfo{pages}{166}
  (\bibinfo{year}{1952}{\natexlab{a}}).

\bibitem[{\citenamefont{Bohm}(1952{\natexlab{b}})}]{bohm:111333}
\bibinfo{author}{\bibfnamefont{D.}~\bibnamefont{Bohm}}, \bibinfo{journal}{Phys.
  Rev.} \textbf{\bibinfo{volume}{85}}, \bibinfo{pages}{180}
  (\bibinfo{year}{1952}{\natexlab{b}}).

\bibitem[{\citenamefont{Bohm et~al.}(19878)\citenamefont{Bohm, Hiley, and
  Kaloyerou}}]{bohm:111444}
\bibinfo{author}{\bibfnamefont{D.}~\bibnamefont{Bohm}},
  \bibinfo{author}{\bibfnamefont{B.~J.} \bibnamefont{Hiley}}, \bibnamefont{and}
  \bibinfo{author}{\bibfnamefont{P.~N.} \bibnamefont{Kaloyerou}},
  \bibinfo{journal}{Phys. Rep.} \textbf{\bibinfo{volume}{144}},
  \bibinfo{pages}{321} (\bibinfo{year}{19878}).

\bibitem[{\citenamefont{Wyatt}(1999)}]{wyatt:111}
\bibinfo{author}{\bibfnamefont{R.~E.} \bibnamefont{Wyatt}},
  \bibinfo{journal}{Chemical Physics Letters} \textbf{\bibinfo{volume}{313}},
  \bibinfo{pages}{189} (\bibinfo{year}{1999}).

\bibitem[{\citenamefont{Garashchuk and Rassolov}(2003)}]{garashchuk:2482}
\bibinfo{author}{\bibfnamefont{S.}~\bibnamefont{Garashchuk}} \bibnamefont{and}
  \bibinfo{author}{\bibfnamefont{V.~A.} \bibnamefont{Rassolov}},
  \bibinfo{journal}{The Journal of Chemical Physics}
  \textbf{\bibinfo{volume}{118}}, \bibinfo{pages}{2482} (\bibinfo{year}{2003}),
  \urlprefix\url{http://link.aip.org/link/?JCP/118/2482/1}.

\bibitem[{\citenamefont{Garashchuk and Rassolov}(2002)}]{garashchuk:111}
\bibinfo{author}{\bibfnamefont{S.}~\bibnamefont{Garashchuk}} \bibnamefont{and}
  \bibinfo{author}{\bibfnamefont{V.~A.} \bibnamefont{Rassolov}},
  \bibinfo{journal}{Chem. Phys. Lett.} \textbf{\bibinfo{volume}{364}},
  \bibinfo{pages}{562} (\bibinfo{year}{2002}).

\bibitem[{\citenamefont{Lopreore and Wyatt}(1999)}]{Lopreore:5190}
\bibinfo{author}{\bibfnamefont{C.}~\bibnamefont{Lopreore}} \bibnamefont{and}
  \bibinfo{author}{\bibfnamefont{R.~E.} \bibnamefont{Wyatt}},
  \bibinfo{journal}{Physical Review Letters} \textbf{\bibinfo{volume}{82}},
  \bibinfo{pages}{5190} (\bibinfo{year}{1999}).

\bibitem[{\citenamefont{Kendrick}(2003)}]{kendrick:5805}
\bibinfo{author}{\bibfnamefont{B.~K.} \bibnamefont{Kendrick}},
  \bibinfo{journal}{The Journal of Chemical Physics}
  \textbf{\bibinfo{volume}{119}}, \bibinfo{pages}{5805} (\bibinfo{year}{2003}),
  \urlprefix\url{http://link.aip.org/link/?JCP/119/5805/1}.

\bibitem[{\citenamefont{Wyatt and Bittner}(2000)}]{bittner:1111}
\bibinfo{author}{\bibfnamefont{R.~E.} \bibnamefont{Wyatt}} \bibnamefont{and}
  \bibinfo{author}{\bibfnamefont{E.~R.} \bibnamefont{Bittner}},
  \bibinfo{journal}{Journal of Chemical Physics}
  \textbf{\bibinfo{volume}{113}}, \bibinfo{pages}{8898} (\bibinfo{year}{2000}).

\bibitem[{\citenamefont{Hughes and Wyatt}(2002)}]{hugh:112}
\bibinfo{author}{\bibfnamefont{K.~H.} \bibnamefont{Hughes}} \bibnamefont{and}
  \bibinfo{author}{\bibfnamefont{R.~E.} \bibnamefont{Wyatt}},
  \bibinfo{journal}{Chemical Physics Letters} \textbf{\bibinfo{volume}{366}},
  \bibinfo{pages}{336} (\bibinfo{year}{2002}).

\bibitem[{\citenamefont{Berry}(1994)}]{berry:111}
\bibinfo{author}{\bibfnamefont{R.~S.} \bibnamefont{Berry}},
  \bibinfo{journal}{J. Phys. Chem} \textbf{\bibinfo{volume}{98}},
  \bibinfo{pages}{6910} (\bibinfo{year}{1994}).

\bibitem[{\citenamefont{Lynden-Bell and Wales}(1994)}]{lynden-bell:1460}
\bibinfo{author}{\bibfnamefont{R.~M.} \bibnamefont{Lynden-Bell}}
  \bibnamefont{and} \bibinfo{author}{\bibfnamefont{D.~J.} \bibnamefont{Wales}},
  \bibinfo{journal}{The Journal of Chemical Physics}
  \textbf{\bibinfo{volume}{101}}, \bibinfo{pages}{1460} (\bibinfo{year}{1994}),
  \urlprefix\url{http://link.aip.org/link/?JCP/101/1460/1}.

\bibitem[{\citenamefont{Predescu et~al.}(2005)\citenamefont{Predescu,
  Frantsuzov, and Mandelshtam}}]{predescu:154305}
\bibinfo{author}{\bibfnamefont{C.}~\bibnamefont{Predescu}},
  \bibinfo{author}{\bibfnamefont{P.~A.} \bibnamefont{Frantsuzov}},
  \bibnamefont{and} \bibinfo{author}{\bibfnamefont{V.~A.}
  \bibnamefont{Mandelshtam}}, \bibinfo{journal}{The Journal of Chemical
  Physics} \textbf{\bibinfo{volume}{122}}, \bibinfo{eid}{154305}
  (pages~\bibinfo{numpages}{12}) (\bibinfo{year}{2005}),
  \urlprefix\url{http://link.aip.org/link/?JCP/122/154305/1}.

\bibitem[{\citenamefont{Chakravarty}(1995{\natexlab{b}})}]{chakravarty:956}
\bibinfo{author}{\bibfnamefont{C.}~\bibnamefont{Chakravarty}},
  \bibinfo{journal}{The Journal of Chemical Physics}
  \textbf{\bibinfo{volume}{102}}, \bibinfo{pages}{956}
  (\bibinfo{year}{1995}{\natexlab{b}}),
  \urlprefix\url{http://link.aip.org/link/?JCP/102/956/1}.

\bibitem[{\citenamefont{Chakravarty}(2002)}]{chakravarty:8938}
\bibinfo{author}{\bibfnamefont{C.}~\bibnamefont{Chakravarty}},
  \bibinfo{journal}{The Journal of Chemical Physics}
  \textbf{\bibinfo{volume}{116}}, \bibinfo{pages}{8938} (\bibinfo{year}{2002}),
  \urlprefix\url{http://link.aip.org/link/?JCP/116/8938/1}.

\bibitem[{\citenamefont{Livesly}(1983)}]{livesly:111}
\bibinfo{author}{\bibfnamefont{D.~M.} \bibnamefont{Livesly}},
  \bibinfo{journal}{J. Phys. C: Solid State Phys.}
  \textbf{\bibinfo{volume}{16}}, \bibinfo{pages}{2889} (\bibinfo{year}{1983}).

\bibitem[{\citenamefont{M.~Toda and Saito}(1992)}]{kubo:111}
\bibinfo{author}{\bibfnamefont{R.~K.} \bibnamefont{M.~Toda}} \bibnamefont{and}
  \bibinfo{author}{\bibfnamefont{N.}~\bibnamefont{Saito}},
  \emph{\bibinfo{title}{Statistical Physics I: Equilibrium Statistical
  Mechanics}} (\bibinfo{publisher}{Berlin:Springer}, \bibinfo{year}{1992}).

\end{thebibliography}

\end{document}